\documentclass[preprint,showpacs,groupedaddress,amsmath,amssymb]{revtex4}
\usepackage{graphicx}
\usepackage{dcolumn}
\usepackage{multirow}
\usepackage{amsfonts}

\newcommand{\beq}{\begin{equation}}
\newcommand{\eeq}{\end{equation}}
\newcommand{\bea}{\begin{eqnarray}}
\newcommand{\eea}{\end{eqnarray}}

\newcommand{\vepsi}{\epsilon}

\newcommand{\ahat}{\hat{a}}
\newcommand{\bhat}{\hat{b}}
\newcommand{\rhat}{\hat{r}}

\newcommand{\thetahat}{\hat{\theta}}

\newcommand{\wt}[1]{\widetilde{#1}}
\newcommand{\mbf}[1]{\mathbf{#1}}

\newcommand{\del}{\nabla}

\newcommand{\curl}{\nabla \times}

\newcommand{\ddt}[1]{\dfrac{\partial\, {#1}}{\partial t}}

\newcommand{\dddtt}[1]{\dfrac{\partial^2\, {#1}}{\partial t^2}}

\newcommand{\oover}[1]{\dfrac{1}{#1}}

\makeatletter
\def\Tbar{\mathchoice
   {\TTbar\displaystyle\textstyle{-}}%
   {\TTbar\textstyle\scriptstyle{-}}%
   {\TTbar\scriptstyle\scriptscriptstyle{-}}%
   {\TTbar\scriptscriptstyle\scriptscriptstyle{-}}%
   \!T}
\def\TTbar#1#2#3{{\setbox0=\hbox{$#1{#2#3}{\mathrm{T}}$}
     \raise2\p@\vbox{\hbox{$#2#3$}}\kern-.35\wd0}}
\makeatother

\begin{document}

\title{Comment on ``Plasma ionization by annularly bounded helicon waves'' [Phys. Plasmas 13, 063501 (2006)]}


\author{Robert W. Johnson}
\email[]{robjohnson@alphawaveresearch.com}
\affiliation{Alphawave Research, Atlanta, GA 30238, USA}

\date{\today}

\begin{abstract}
The neoclassical calculation of the helicon wave theory contains a fundamental flaw.  Use is made of a proportional relationship between the magnetic field and its curl to derive the Helmholtz equation describing helicon wave propagation; however, by the fundamental theorem of Stokes, the curl of the magnetic field must be perpendicular to that portion of the field contributing to the local curl.  Reexamination of the equations of motion indicates that only electromagnetic waves propagate through a stationary region of constant pressure in a fully ionized, neutral medium.
\end{abstract}

\pacs{52.25.Jm, 52.50.Dg, 52.40.Fd} 

\maketitle


The neoclassical calculation of the helicon wave theory, presented for the case of an annular waveguide in Reference~\cite{yano-063501} which follows Chen's derivation~\cite{chen-137} for a cylindrical configuration, contains a fundamental flaw in its use of vector field theory~\cite{griffiths-89,ryder-qft}.  A proportional relationship is derived between the magnetic field and its curl which contradicts the fundamental theorem of Stokes, thus the waves described cannot be physical.  Reexamination of the plasma equations of motion with consideration of the convective terms for the case of vanishing mass flow, charge density, and pressure gradients leads to the familiar Ohm's law of conductive material, indicating that only the usual electromagnetic radiation propagates through such regions.  A similar discussion on the existence of whistler oscillitons~\cite{npg-12-425-2005,npg-14-49-2007,npg-14-543-2007,npg-14-545-2007} in geophysical plasmas is noted.

\begin{figure}
\includegraphics[scale=.5]{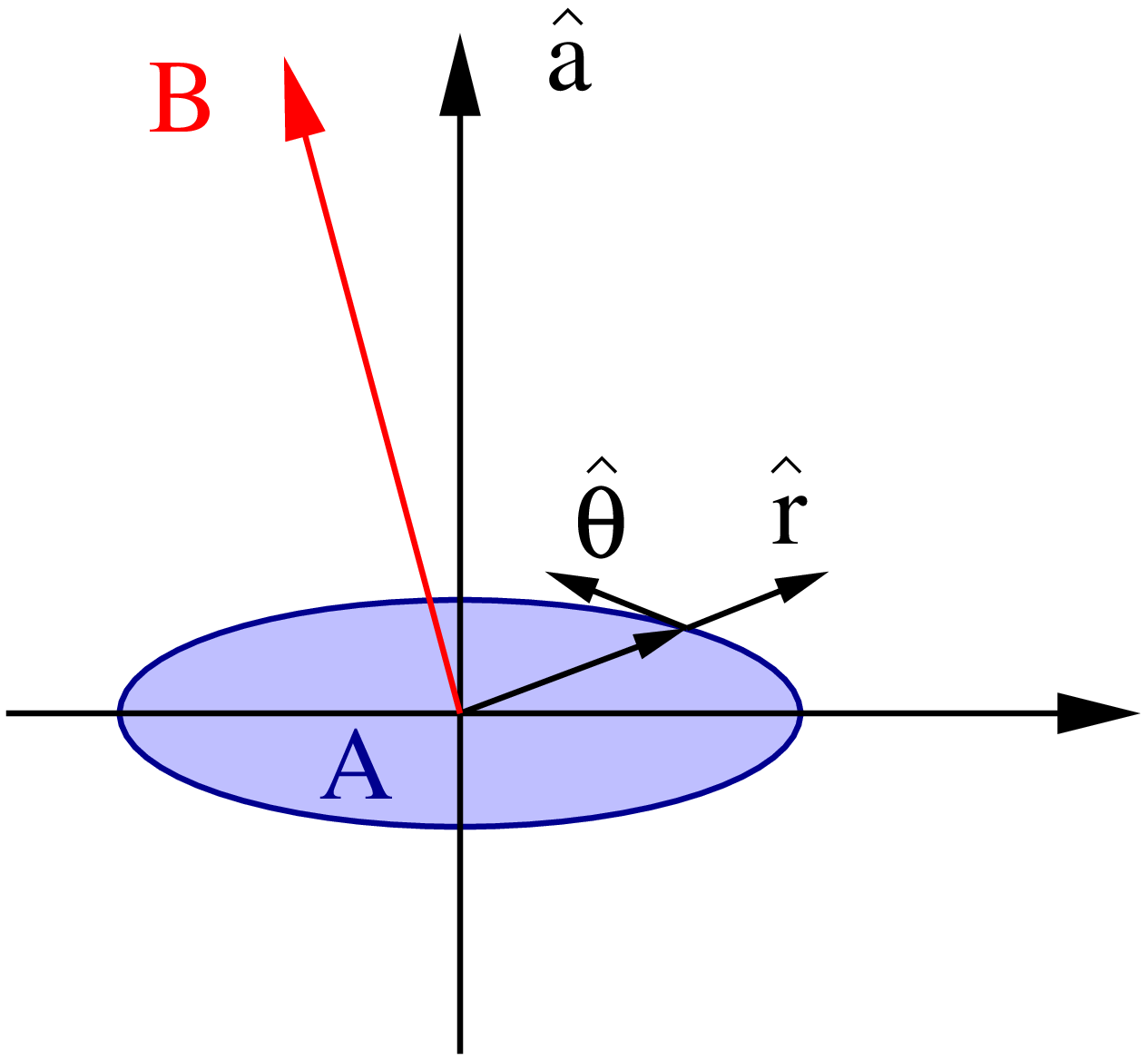}%
\caption{\label{fig-curl}(Color online.)  The definition of the curl of a vector field is the circulation per area, $(\curl \mbf{B}) \cdot \ahat \equiv \lim_{A \rightarrow 0} \left[\oint_{\partial A} dl \ \mbf{B}(\mbf{r}) \cdot (\ahat \times \rhat) \right]/A$.  The curl of a real vector field must everywhere be orthogonal to the local plane of circulation, hence that portion of the vector contributing to the curl at that location.}
\end{figure}

The model for infinite conductivity $\sigma$ is given by the linearized system: \beq
\curl \mbf{E} = - \partial \mbf{B} / \partial t \;,\;\; \curl \mbf{B} = \mu_0 \mbf{J} \;,\;\; \mbf{E} = \mbf{J} \times \mbf{B}_0 / e n_0 \;,
\eeq where $\mbf{B}_0$ is the applied magnetic field along some axis $\bhat$ and $n_0$ is the plasma density $n_e = n_i = n_0$, with solutions of the form $\wt{\mbf{B}} = \mbf{B}(r) e^{i(m \theta + k z - \omega t)}$.  The standard derivation of the helicon wave equation is here termed ``neoclassical'' for its use of the quasineutral approximation to justify its determination of the electric field from an equation of motion rather than the remainder of Maxwell's equations~\cite{maxwell-1864}, as given by the substitution apparent in Equation~(9) of Reference~\cite{yano-063501}, \beq \label{eqn-bandj}
i \omega \wt{\mbf{B}} = \curl \wt{\mbf{E}} = \curl (\wt{\mbf{J}} \times \mbf{B}_0)/e n_0 = (\mbf{B}_0 \cdot \del) \wt{\mbf{J}}/e n_0 = \wt{\mbf{J}} \left(i k B_0/e n_0\right) \;,
\eeq where we note that implicit use of a constant density has been made, as \beq
\curl (\mbf{J} \times \mbf{B}_0 / n_0) = [\curl (\mbf{J} \times \mbf{B}_0)]/n_0 - (\mbf{J} \times \mbf{B}_0) \times \del (1/n_0) \;,
\eeq with the result \beq \label{eqn-alphaB}
\curl \wt{\mbf{B}} = \alpha \wt{\mbf{B}} \;,
\eeq for $\alpha = (\omega/k)(\mu_0 e n_0 / B_0) = (\omega/k)(\omega_{pe}^2/\omega_{ce} c^2)$, leading to the Helmholtz wave equation $(\del^2 + \alpha^2) \wt{\mbf{B}} = 0$.  However, for the real field $\mbf{B}$ displaying circulation caused by the local current $\mbf{J}$ which from Equation~(\ref{eqn-bandj}) is in phase, by the fundamental theorem of Stokes~\cite{griffiths-89,wolf-curl}, \beq
\left(\curl \mbf{B}\right) \cdot \bhat \equiv \lim_{A \rightarrow 0} \dfrac{\oint_{\partial A} dl \ \mbf{B}(\mbf{r}) \cdot (\bhat \times \rhat)}{A} = \lim_{r \rightarrow 0} \dfrac{\oint_0^{2 \pi} r d\theta \ \mbf{B(\mbf{r})} \cdot \thetahat}{\pi r^2} = 0 \;,
\eeq where these $r$ and $\theta$ refer to our local geometry, not the global one, as in Figure~\ref{fig-curl}, the component of the curl of $\mbf{B}$ along $\mbf{B}$ must vanish, thus Equation~(\ref{eqn-alphaB}) cannot hold for both $\alpha$ and $\mbf{B}$ nontrivially real.  As the wave in question propagates freely without attenuation, $k$ hence $\alpha$ is real, leading to an infinite wave number for finite $\omega$ or a zero frequency for finite $k$.  One might patch the theory by inclusion of the displacement term in the Maxwell-Ampere equation, necessary when the rate of change of $\mbf{B}$ is itself changing, $\partial^2 \mbf{B} / \partial t^2 \neq 0$, and essential to the propagation of electromagnetic waves, so that the equation $\curl (\curl \mbf{B}) = \mu_0 \curl (\mbf{J} + \vepsi_0 \partial \mbf{E} / \partial t)$ leads to \beq \label{eqn-KGB}
\del^2 \mbf{B} + \alpha \left( \alpha \mbf{B} + \mu_0 \vepsi_0 \ddt{\mbf{E}} \right) - \mu_o \vepsi_0 \dddtt{\mbf{B}} = \left[ \Box^2 + \alpha^2 \left( 1 - \dfrac{\vepsi_0}{e n_0} \mbf{B}_0 \times \ddt{} \right) \right] \mbf{B} = 0 \;,
\eeq where we recognize a modified Klein-Gordon equation for each component of $\mbf{B}$, admitting \beq
\left[ \del^2 + \alpha^2 + \mu_0 \vepsi_0 \omega^2 \left( 1 + i \dfrac{\alpha}{k} \bhat_0 \times \right) \right] \wt{\mbf{B}} = 0 \;.
\eeq  For consistency, one should require of Equation~(\ref{eqn-KGB}) reduction to the standard wave equation for conductors $\Box^2 \mbf{B} = \mu_0 \sigma \partial \mbf{B} / \partial t$ in the limit of vanishing applied magnetic field $\mbf{B}_0 \rightarrow 0$, which admits attenuated solutions with finite $\omega$ and $k$.  The first problem to arise is that $\lim_{B_0 \rightarrow 0} \alpha = \infty$ rather than 0, and the second is that an infinite conductivity $\sigma \rightarrow \infty$ requires $\partial \mbf{B} / \partial t \rightarrow 0$, hence a static magnetic field.

The difficulty encountered above by the neoclassical model results from the misapplication of the Ohm's law equation, which itself is derived from the electron and ion equations of motion.  For a neutral, hydrogenic plasma of species $s\in\{e,i\}$ with total particle density $n \equiv n_e+n_i=2 n_0$, mass density $\rho_m \equiv \sum_s n_s m_s$, mass flow velocity $\rho_m \mbf{V}_m \equiv \sum_s n_s m_s \mbf{V}_s$, free current density $\mbf{J}_f \equiv \sum_s n_s e_s \mbf{V}_s$, and pressure $p \equiv n\, \Tbar \equiv \sum_s n_s \Tbar_s \equiv \sum_s p_s$, the equations of motion neglecting viscosity and gyromotion read \beq \label{eqn:eqnofmtn}
n_s m_s \left[\ddt{} + \left( \mbf{V}_s \cdot \del \right) \right] \mbf{V}_s + \del p_s = n_s e_s \left( \mbf{E} + \mbf{V}_s \times \mbf{B} \right) + \mbf{F}_{s k}\;,
\eeq where $k\neq s$ for the friction term $\mbf{F}_{e i} = - \mbf{F}_{i e} = m_e \nu_{e i} \mbf{J}_f/e$ which represents interspecies collisions.  From the definitions of the current and flow velocity we may exchange the electron and ion velocities $\{\mbf{V}_e, \mbf{V}_i\}$ for the pair $\{\mbf{V}_m, \mbf{J}_f\}$ via \beq
\left[\begin{array}{c} \rho_m \mbf{V}_m\\\mbf{J}_f \end{array} \right] = n_0 \left[\begin{array}{cc} m_e & m_i \\ -e & e \end{array} \right] \left[\begin{array}{c} \mbf{V}_e\\\mbf{V}_i \end{array} \right] \Rightarrow \left[\begin{array}{c} \mbf{V}_e\\\mbf{V}_i \end{array} \right] = \mbf{V}_m + \dfrac{\mbf{J}_f}{e \rho_m} \left[\begin{array}{r} \ -m_i\\m_e \end{array} \right]\;.
\eeq  The sum of Equations~(\ref{eqn:eqnofmtn}) gives the net force balance equation \beq \label{eqn:equileqn}
\rho_m \ddt{\mbf{V}_m} + \mbf{C}_+ + \del p = \mbf{J}_f \times \mbf{B} \;,
\eeq and their difference the generalized Ohm's law equation \beq \label{eqn:ohmseqn}
n_0 (m_i - m_e) \ddt{\mbf{V}_m} + \dfrac{2 m_e m_i}{e (m_i + m_e)} \ddt{\mbf{J}_f} + \mbf{C}_- + \del \left( p_i - p_e \right) = n_0 e \left( \mbf{V}_i + \mbf{V}_e \right) \times \mbf{B} + 2 \left[ n_0 e \mbf{E} - \mbf{F}_{e i} \right]\;,
\eeq where $n_0 e (\mbf{V}_i+\mbf{V}_e) = 2 n_0 e \mbf{V}_m - n_0 (m_i-m_e) \mbf{J}_f/\rho_m $, and the convective terms $\mbf{C}_{+,-} \equiv n_0 \left[ m_i \left( \mbf{V}_i \cdot \del \right) \mbf{V}_i \pm m_e \left( \mbf{V}_e \cdot \del \right) \mbf{V}_e \right]$ are given by \bea \label{eqn:convterms}
\mbf{C}_+ =& n_0 \left(m_i + m_e\right) & \left[ \left( \mbf{V}_m \cdot \del \right) \mbf{V}_m + \dfrac{m_e m_i}{e^2} \left(\dfrac{\mbf{J}_f}{\rho_m} \cdot \del \right) \dfrac{\mbf{J}_f}{\rho_m} \right] ,\\
\mbf{C}_- =& n_0 \left(m_i - m_e\right) & \left[ \left( \mbf{V}_m \cdot \del \right) \mbf{V}_m - \dfrac{m_e m_i}{e^2} \left(\dfrac{\mbf{J}_f}{\rho_m} \cdot \del \right) \dfrac{\mbf{J}_f}{\rho_m} \right] \\ & + \dfrac{2 n_0 m_e m_i}{e} & \left[ \left( \mbf{V}_m \cdot \del \right) \dfrac{\mbf{J}_f}{\rho_m} + \left(\dfrac{\mbf{J}_f}{\rho_m} \cdot \del \right) \mbf{V}_m \right] \;.
\eea  With consideration now of a vanishing flow velocity $\mbf{V}_m = 0$, the generalized Ohm's law equation may be put into the form \beq \label{eqn:myohm1}
\dfrac{m_e m_i}{e^2 \rho_m} \ddt{\mbf{J}_f} + \dfrac{\mbf{C}_-}{2 n_0 e} + \dfrac{(m_i - m_e)}{2 e \rho_m} \left(\mbf{J}_f \times \mbf{B}\right) = \mbf{E} -  \eta \mbf{J}_f \;,
\eeq for resistivity $\eta=m_e \nu_{e i} / n_0 e^2$ where $\nu_{e i}$ is the interspecies collision rate, when the electrons and ions have the same thermal energy $\Tbar_e = \Tbar_i = \Tbar_0$.  As the net force balance equation must also hold true, for a region of constant pressure $\del p = 0$ we find that $\mbf{J}_f \times \mbf{B} = \mbf{C}_+$.  Then follows \beq
\dfrac{m_e m_i}{e^2 \rho_m} \ddt{\mbf{J}_f} + \oover{2 n_0 e} \left[\mbf{C}_- + \dfrac{n_0 (m_i-m_e)}{\rho_m} \mbf{C}_+ \right] = \dfrac{m_e m_i}{e^2 \rho_m} \ddt{\mbf{J}_f} = \mbf{E} -  \eta \mbf{J}_f \;,
\eeq which, in terms of the conductivity $\sigma \equiv 1 / \eta$ and transit time $\tau_{e i} \equiv 1 / \nu_{e i}$, reduces to \beq
\dfrac{m_i}{(m_i+m_e)} \tau_{e i} \ddt{\mbf{J}_f} = \sigma \mbf{E} - \mbf{J}_f \;.
\eeq  With an oscillating field $\wt{\mbf{E}} = \mbf{E} e^{-i \omega t}$ and current $\wt{\mbf{J}}_f = \mbf{J}_f e^{-i \omega t}$, one finds \beq
\dfrac{m_i}{(m_i+m_e)} \tau_{e i} (- i \omega) \wt{\mbf{J}}_f = \sigma \wt{\mbf{E}} - \wt{\mbf{J}}_f \;,
\eeq and bringing the factors of $\wt{\mbf{J}}_f$ together gives us \beq
\wt{\mbf{J}}_f = \sigma \left[1 - i \omega \tau_{e i} \dfrac{m_i}{(m_i+m_e)} \right]^{-1} \wt{\mbf{E}} \;,
\eeq which we compare to the usual complex conductivity of a material conductor~\cite{griffiths-89}, \beq
\wt{\mbf{J}}_f = \left(\dfrac{n_0 e^2 / m_e}{\nu - i \omega} \right) \wt{\mbf{E}} \;,
\eeq noting that Ohm's law determines $\mbf{J}_f$ in terms of $\mbf{E}$ for a plasma just as in a normal conductor.  With substitution into the Maxwell equations, one achieves the usual modified wave equations for a dispersive conducting medium, leading to the attenuated propagation of electromagnetic radiation, and with vanishing damping factor a plasma frequency of $\omega_p = e \sqrt{n_0 (m_i + m_e)/m_i m_e \vepsi_o}$.

So, where have all the waves gone?  They have disappeared along with the variations in charge, mass, energy, and momentum density given by our assumptions of applicability.  Without a model allowing for the explicit determination of these quantities, one is left with a theory of only electromagnetic waves.  If one wishes to describe the behavior of other types of waves in the plasma, one is forced to deal with the physics describing the oscillation of the plasma medium.  We note that the model above may be extended to incorporate the effects of the gyromotion through the inclusion of the bound current $\mbf{J}_b \equiv \curl \mbf{M}$ for $\mbf{M} \equiv \sum_s n_s \vec{\mu}_s$ and the drift momentum $\rho_m \mbf{V}_d \equiv \curl \mbf{L}_g$ for $\mbf{L}_g \equiv \sum_s n_s \vec{l}_s$, where $\vec{\mu}_s$ and $\vec{l}_s$ are the magnetic moment and angular momentum per particle, respectively, to give a total current $\mbf{J} = \mbf{J}_f + \mbf{J}_b$ and momentum density $\rho_m \mbf{V} = \rho_m (\mbf{V}_m + \mbf{V}_d)$.  As helicon antenna devices are in operation~\cite{aiaa08-4926,ispc08-0236} as prototypes for a high-efficiency ion source and for an electric propulsion thruster, a better determination of what is happening within the waveguide is warranted.



\newpage


\newpage

\end{document}